\def\ra{\rangle}
\def\la{\langle}
\def\be{\begin{equation}}
\def\ee{\end{equation}}
\def\ba{\begin{array}}
\def\ea{\end{array}}

\documentclass[pra,showpacs,amsmath,twocolumn,aps]{revtex4-1}

\def\qed{\leavevmode\unskip\penalty9999 \hbox{}\nobreak\hfill
     \quad\hbox{\leavevmode  \hbox to.77778em{%
               \hfil\vrule   \vbox to.675em%
               {\hrule width.6em\vfil\hrule}\vrule\hfil}}
     \par\vskip3pt}

\begin{document}

\title{Complete Entanglement Witness for Quantum Teleportation}
\author{Ming-Jing Zhao$^{1}$}
\author{Shao-Ming Fei$^{1,2}$}
\author{Xianqing Li-Jost$^{1}$}
\affiliation{$^1$Max-Planck-Institute for Mathematics in the Sciences, 04103
Leipzig, Germany\\
$^2$School of Mathematical Sciences, Capital Normal
University, Beijing 100048, China}

\begin{abstract}
We propose a set of linear quantum entanglement witnesses constituted by local quantum-mechanical
observables with each two possible measurement outcomes. These witnesses detect all the entangled resources
which give rise to a better fidelity than separable states in quantum teleportation and present both sufficient and
necessary conditions in experimentally detecting the useful resources for quantum teleportation.
\end{abstract}

\maketitle

{\it Introduction.}~ Quantum entanglement plays important roles in many quantum information
processing such as quantum teleportation.
However, not all entangled states are useful for quantum teleportation.
The fidelity of optimal teleportation is determined by the fully
entangled fraction (FEF) \cite{bennett,M. Horodecki1999,alb2002}.
A bipartite $n\otimes n$ state $\rho$ gives rise to a better fidelity of teleportation
than separable states if its FEF is great than $1/n$.
For a known quantum state, analytical formula of FEF for two-qubit states has been
derived by using the method of Lagrange multiplier \cite{grondalski}.
The upper bounds of FEF for general high dimensional quantum states have been estimated \cite{upperbound}.
Exact results of FEF are also obtained for some special quantum states like isotropic states
and Werner states \cite{mj2010}.

For a given unknown state, an important issue is to
determine whether it is useful for quantum teleportation by experimental measurements.
For the experimental detection of quantum entanglement,
the Bell inequalities \cite{bell1,bell2,bell3,bell4,liprl} and
entanglement witness \cite{5,9,11,12,13,15,wit4,wit5,yusx,zhaomj}
have been extensively investigated.
The Bell inequalities only involve measurements on local quantum mechanical observables.
The entanglement witnesses are in general
Hermitian operators with at least one negative eigenvalue.
Since the set of the separable states is convex and compact,
these witnesses give rise to inequalities (supersurfaces) separating a part of the
entangled states from the rest ones
including all the separable states. Different inequalities detect different entangled states.
However, so far we do not have complete witnesses that detect all the entangled states in general.

Recently in Ref. \cite{N. Ganguly}, the authors show that the set of entangled states which are useful
for quantum teleportation, i.e. their FEFs are great than $1/n$, is also convex and compact.
They presented a witness operator which detects some entangled states that are useful for teleportation.

In this brief we give a general way to enquire how to
determine experimentally whether an unknown entangled state could be
used as a resource for quantum teleportation.
We present a linear witness operator which can detect all the entangled states that are useful
for teleportation. This witness operator gives rise to a Bell-like inequality which
requires only measurements on local observables and gives the sufficient and necessary condition
for states that are useful for teleportation.

\smallskip
{\it Linear witness for quantum teleportation.}~ Let $H_n$
be an $n$-dimensional complex Hilbert space,
with $\{\vert i\rangle\}_{i=1}^n$ an orthonormal basis.
Let $\rho$ be a density matrix defined on $H_n\otimes H_n$.
The optimal fidelity of teleportation with $\rho$ as the entangled resource is given by \cite{bennett,M. Horodecki1999,alb2002}
\be\label{f}
f_{max}(\rho)=\frac{n F(\rho)}{n+1}+\frac{1}{n+1}.
\ee
$F(\rho)$ is the fully entangled fraction with respect to $\rho$:
\begin{eqnarray}\label{FEF}
F(\rho)=\max_U \langle \psi^+| (U^\dagger \otimes I_n)\, \rho\, (U \otimes I_n) |\psi^+\rangle,
\end{eqnarray}
where $U$ is any $n\times n $ unitary matrix, $I_n$ is the $n\times n$ identity matrix, and $|\psi^+\rangle$
is the maximally entangled state,
$$
|\psi^+\rangle=\frac{1}{\sqrt{n}}\sum_{i=0}^{n-1}|ii\rangle.
$$

A state $\rho$ is a useful resource for teleportation if and only if $F(\rho)>\frac{1}{n}$ \cite{M. Horodecki1999}.
If $F(\rho)\leq\frac{1}{n}$, the fidelity (\ref{f}) is no better than separable states.
Although (\ref{FEF}) has no analytical formula for a general $\rho$ so far, in the following we show
that one can detect completely that if $F(\rho)>\frac{1}{n}$ by experimental measurements on local observables.

Let us define the Hermitian operators,
$$\ba{l}
\lambda_i=|0\rangle\langle0| - |i\rangle\langle i|,\\[2mm]
\lambda_{kl}=|k\rangle\langle l| + |l\rangle\langle k|,\\[2mm]
\lambda_{kl}^\prime={\rm i}(|k\rangle\langle l| - |l\rangle\langle k|),
\ea
$$
with $k<l$, and $i,k,l=1,\cdots,n-1$. Set $A_i= U\lambda_i U^\dagger$, $A_{kl}=U \lambda_{kl} U^\dagger$, $A_{kl}^\prime=U \lambda_{kl}^\prime U^\dagger$, with $U$ any $n\times n$ unitary matrix.
We define the linear witness operator to be
\be\label{w}
\begin{array}{rcl}
\displaystyle \Gamma&\equiv& \displaystyle \frac{1}{n}[ I_n \otimes I_n
+n\sum_{i=1}^{n-1}A_i\otimes \lambda_i - \sum_{i=1}^{n-1}\sum_{j=1}^{n-1}A_i\otimes \lambda_j]\\[2mm]
&&\displaystyle +\frac{1}{2}\sum_{k< l}( A_{kl} \otimes \lambda_{kl} -A_{kl}^\prime \otimes\lambda_{kl}^\prime).
\end{array}
\end{equation}
From the mean value of $\Gamma$, $\langle \Gamma\rangle_{\rho}=Tr (\Gamma\rho)$, with $Tr$ standing
for trace, we have

\medskip
{\bf Theorem}~
$\rho$ is useful for teleportation if and only if the mean value of $\Gamma$ satisfies,
\be\label{thm}
\langle \Gamma\rangle_{\rho}>1.
\ee

\medskip
{[Proof]}. By expanding the operator $|\psi^+\rangle \langle \psi^+|$ according to the Hermitian operators
$\lambda_i$, $\lambda_{kl}$ and $\lambda_{kl}^\prime$, we have
$$
\begin{array}{l}
\displaystyle n|\psi^+\rangle \langle \psi^+|\\[2mm]
=\displaystyle \frac{1}{n}[ I_n \otimes I_n
+n\sum_{i=1}^{n-1}\lambda_i\otimes \lambda_i - \sum_{i=1}^{n-1}\sum_{j=1}^{n-1}\lambda_i\otimes \lambda_j]\\[2mm]
~~\displaystyle +\frac{1}{2}\sum_{k< l}( \lambda_{kl} \otimes \lambda_{kl} -\lambda_{kl}^\prime \otimes\lambda_{kl}^\prime),
\end{array}
$$
i.e. $\Gamma=n (U \otimes I_n)|\psi^+\rangle \langle \psi^+|(U^\dagger \otimes I_n)$.
Therefore we have the following relation,
$$
\ba{rcl}
\langle \Gamma\rangle_{\rho}&=&\langle n (U \otimes I_n)|\psi^+\rangle \langle \psi^+|(U^\dagger \otimes I_n)\rangle_{\rho}\\[2mm]
&=&n \langle \psi^+| (U^\dagger \otimes I_n)\, \rho\, (U \otimes I_n) |\psi^+\rangle
\ea
$$
and
$$
n F(\rho)=\max_U \langle \Gamma\rangle_{\rho}.
$$
Hence $\rho$ is useful for quantum teleportation, $F(\rho)>\frac{1}{n}$, if and only if $\langle \Gamma\rangle_{\rho}>1$.
\qed

This implies that these linear operators $\{\Gamma\}$ have the ability to witness all the useful resources for quantum teleportation. In other words, the supersurfaces given by the inequality (\ref{thm}) of
these linear operators separate geometrically all the useful resources of quantum teleportation from the convex and compact set of all states that are not useful for quantum teleportation. This fact distinguishes the FEF from the entanglement measures for which there have been no complete set of entanglement witnesses so far that detect all the entangled states.

\smallskip
{\it Particular cases}.~ Let us consider the two-qubit case. From (\ref{thm})
we get that $\rho$ is useful for quantum teleportation if and only if
\begin{eqnarray}\label{2by2}
\langle A_z\otimes \sigma_z +A_x\otimes \sigma_x-A_y\otimes \sigma_y\rangle_{\rho}>1,
\end{eqnarray}
where $\sigma_z=|0\rangle\langle0|-|1\rangle\langle1|$, $\sigma_x=|0\rangle\langle1|+|1\rangle\langle0|$, and $\sigma_y={\rm i}(|0\rangle\langle1|-|1\rangle\langle0|)$ are Pauli matrices, $A_z=U\sigma_zU^\dagger$, $A_x=U\sigma_xU^\dagger$ and $A_y=U\sigma_yU^\dagger$.

One may compare this situation to entanglement detection. In fact for two-qubit case, all entangled states can be also detected experimentally. For pure two-qubit states, the CHSH-Bell inequality \cite{bell3},
$|\la A_1\otimes B_1+A_1\otimes B_2+A_2\otimes B_1-A_2\otimes B_2 \ra|\leq2$, gives
a sufficient and necessary condition of separability, where
$A_i=\vec{a_i}\cdot \vec{\sigma}_A$,
$B_j=\vec{b_j}\cdot \vec{\sigma}_B$,
$\vec{a_i}=(a_i^x,a_i^y,a_i^z)$ and $\vec{b_j}=(b_j^x,b_j^y,b_j^z)$
are real unit vectors satisfying $|\vec{a_i}|=|\vec{b_j}|=1$, $i,j=1,2$, $\vec{\sigma}_{A/B}=(\sigma^{A/B}_{x},\sigma^{A/B}_{y},\sigma^{A/B}_{z})$,
$\sigma^{A/B}_{x,y,z}$ are Pauli matrices associated to the qubit $A$ and $B$. Here to detect
the entanglement one needs to measure four observables along all possible directions of spin.
While from inequality (\ref{2by2}), to detect the teleportion resource one only needs to
measure three observables of one qubit under all $U$, and just fixed observables
$\sigma_{x,y,z}$ of another qubit. Moreover, inequality (\ref{2by2}) is valid also for
mixed states, for which the entanglement detection is given by a more complicated
non-linear inequality \cite{yusx}.

As for a detailed example, we consider a state which can not be detected by the witness
presented in Ref. \cite{N. Ganguly},
\begin{eqnarray}\label{ex}
\rho=a|\phi\rangle \langle \phi|+(1-a)|11\rangle\langle11|,
\end{eqnarray}
where $|\phi\rangle=\frac{1}{\sqrt{2}}(|01\rangle +|10\rangle)$ and $0\leq a\leq 1$.
The usefulness of the state (\ref{ex}) as a teleportation resource can be detected by our
inequality (\ref{2by2}). In fact, we can simply chose
$U=|0\rangle\langle1|+|0\rangle\langle1|$. Then the difference between the left hand side and the right hand side of inequality (\ref{2by2}) is
$\langle A_z\otimes \sigma_z +A_x\otimes \sigma_x-A_y\otimes \sigma_y\rangle_{\rho}-1=
4a-2$, which is positive if and only if $a>\frac{1}{2}$.
Therefore our witness operator can completely detect the usefulness of $\rho$ in quantum teleportation.

{\it Conclusion.}~
We have provided a complete set of linear witness operators which gives rise to sufficient and necessary conditions in experimentally detecting the useful resources for quantum teleportation. These linear witnesses are composed of only local observables with two possible outcomes each.

\end{document}